# Magnetically tunable electrokinetic instability and structuring of non-equilibrium nanoparticle gradients

F. Sohrabi,[†1] C. Rigoni,[†1] T. Cherian,[1] O. Ikkala,[1] J. V. I. Timonen,[1]*

[1]Department of Applied Physics, Aalto University School of Science,

Puumiehenkuja 2, 02150 Espoo, Finland

*Corresponding author; E-mail: jaakko.timonen@aalto.fi

[†] Equal contribution

## Abstract

Inspired by emergent behaviors of living matter, there is increasing interest in developing approaches to create dynamic patterns and structures in synthetic materials with controllable complexity to enable functionalities that are not possible in thermodynamic equilibrium. Here we show that electrophoretically driven and maintained non-equilibrium gradients of magnetic nanoparticles in non-polar solvent can undergo electrokinetic instabilities (EKI), leading to various electrically controllable spatiotemporally patterned states. These electrokinetic instabilities and patterns can be tuned with a magnetic field via magnetostatic energy reduction mechanism to both increase and decrease the pattern complexity. We reflect the experimental observations on the theoretical electrokinetic and magnetostatic arguments. We further show that small amounts of polar water in the otherwise non-polar system are critical enablers for the electrophoretic mobility of the nanoparticles. Since functionalities of magnetic nanoparticles are widely tunable, we foresee that the combination of dissipative electrokinetic driving and magnetic energy reduction can lead to novel functional dissipative materials.

# I. INTRODUCTION

Living materials exhibit exceptional non-equilibrium structures critical for life [1]. A fundamental challenge in modern materials science is how to find approaches to produce analogous structures in simple synthetic materials to create life-like functionalities. However, in contrast to widely studied equilibrium patterns and structures, that can be rationally designed and produced in synthetic materials [2–4], the dissipative non-equilibrium patterns require continuous entropy production and are considerably more complicated to realize [5–9].

One powerful approach to drive non-equilibrium states in soft materials like colloids is to use electric fields [10–14]. Application of uniform linear [15,16] or rotating/biaxial [17,18] AC fields can be used to alter interparticle interactions, leading to e.g. formation of ordered lattices or zigzag-like assemblies [19]. If the field is not uniform, the additional dielectrophoretic force can drive the particles into spatially varying concentration gradients [20–22] that can be considered as an elementary perturbation of the uniform distribution of the thermodynamic ground state. Such material gradients can be driven, in some cases, even using uniform DC fields [23–26]. Interestingly, such material gradients in electric fields are often unstable and undergo electrokinetic instabilities (EKI) [27]. These instabilities have been studied both theoretically [28–30] and often experimentally in microfluidic settings [31–33] where the gradient is not driven electrically but maintained independently using fluid flow. While the EKIs are particularly interesting for micromixing in lab-on-chip applications [34,35], the resulting material flows are often complex and chaotic.

One approach to tuning instabilities is the use of magnetic fields and magnetic fluids (ferrofluids). Previously ferrofluids have been used to tune and control the properties of hydrodynamic instabilities like turbulent waves [36,37] oscillating droplets [38] and for the Kelvin-Helmholtz instability [39]. They are also interesting from pattern formation perspective as they exhibit pattern formation in external magnetic field [40], including the Rosensweig instability [2] and the labyrinthine instability [41] and their analogies on a diffuse electrically driven interface [26]. The effect of electric fields on ferrofluids has been explored to limited extent, and control over susceptibility [42,43] and electrokinetic/electroconvective instabilities in microfluidic channels [44–46], bulk volumes in glass cuvettes [47] and in Hele-Shaw cells [48–52] have been demonstrated. However, surprisingly magnetic control of these phenomena has explored only to limited extent [50,51] to demonstrate how magnetic fields can be used to tune the onset and the features of electroconvection patterns.

In this work, we demonstrate well-defined, magnetically controllable electrokinetic instabilities in a dispersion of magnetic iron oxide nanoparticles in dodecane, whose electrical response is accurately tunable by addition of charge carrying surfactant reverse micelles. Firstly, we start by demonstrating that the electric responsivity and formation of nanoparticle gradients is strongly dependent on the amount of trace water in this nominally nonpolar system. Secondly, we show that these elementary and reversible non-equilibrium gradients can undergo various electrokinetic instabilities and patterning, which we address to space charge build-up in within the gradient. Thirdly, we show how these electrokinetically driven patterns can be tuned by application of a magnetic field to create patterns and structures of broadly tunable complexity.

## II. RESULTS

### A. Overview of the experimental system

The electrically and magnetically responsive sample was prepared as described earlier by combining superparamagnetic iron oxide nanoparticles stabilized with oleic acid and inverse micelles of AOT (bis(2-ethylhexyl) sulfosuccinate sodium salt) in dodecane [Fig. 1a,b]. The magnetic responsivity of the electroferrofluid is bounded to the concentration of magnetic nanoparticles while the electrical responsivity is, in addition, also dependent on the amount of AOT and trace water [Fig. 1b]. Through application of electric field, it is possible to drive the sample outside the equilibrium and obtain electrokinetic states with different properties, application of magnetic fields can further tune these electrokinetic states by acting on the minimization of magnetic energy [Fig. 1c]. Electrokinetic driving and magnetic energy minimization both act on changing the spatial distribution of the nanoparticles, hence the state of the system that occurs upon application of both fields is balanced by the constant competition of the two phenomena [Fig. 1d]. In our experiments the colloid was subjected to electric field by using a microelectrode cell with planar interdigitated indium tin oxide (ITO) electrodes [Fig. 1e] that was further placed in uniform magnetic field generated with Helmholtz coils. As a response to electric and magnetic fields, the sample transitions from a uniform dispersion (thermodynamic ground state) [Fig. 1f] to various non-equilibrium states that were explored in detail: stationary nanoparticle gradients driven by electrophoresis, magnetically patterned stationary gradients, electrokinetically driven spatiotemporal patterns, and magnetically tuned electrokinetically driven spatiotemporal patterned states [Fig. 1g].

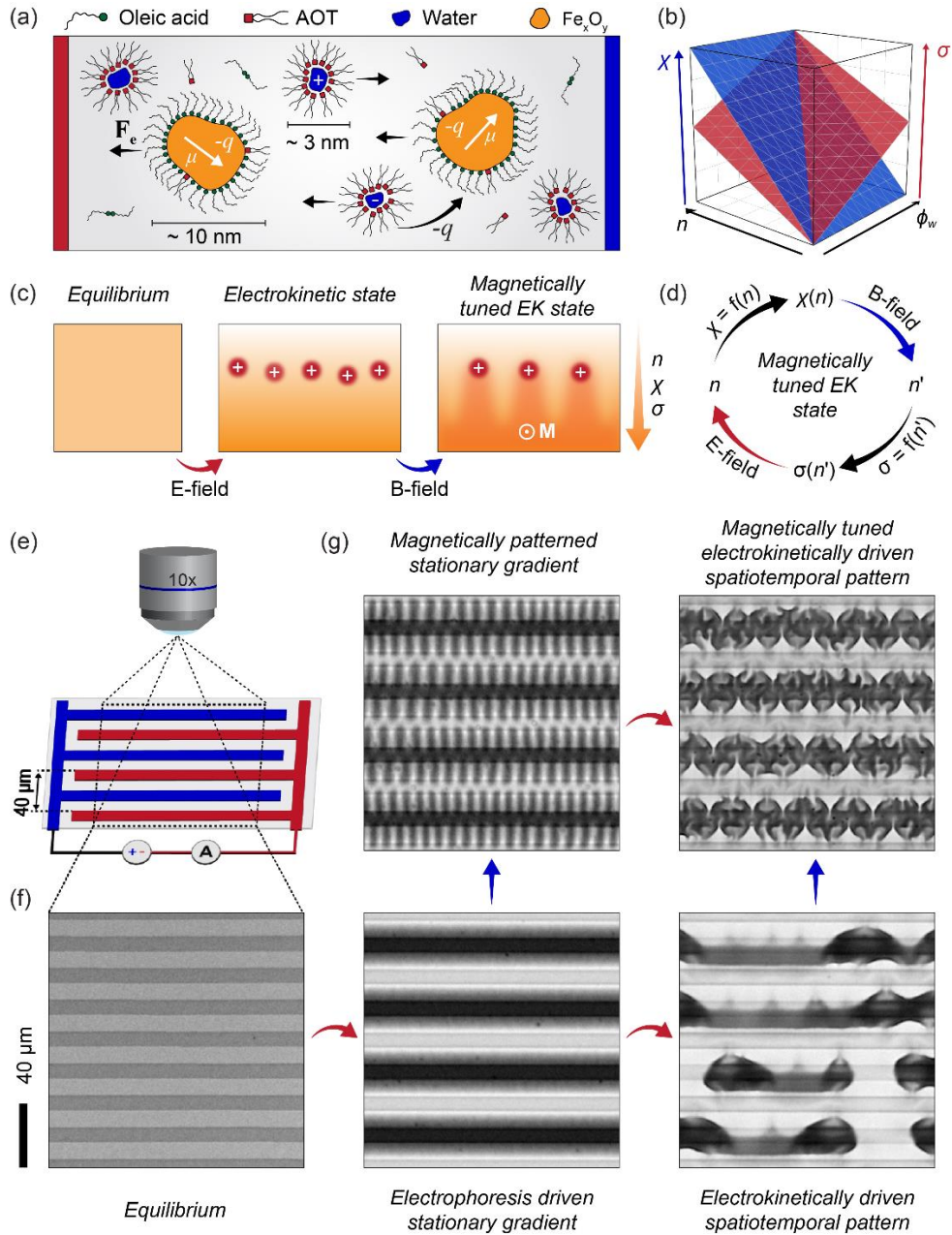

FIG. 1. Overview and demonstration of electrokinetically driven and magnetically tunable patterning of nanoparticle gradients. (a) A scheme of the composition of the electrically and magnetically responsive colloidal dispersion used in the experiments. (b) Diagram of the dependency of magnetic susceptibility ($\chi$) and electric conductivity ($\sigma$) as a function of the nanoparticle concentration ($n$) and the water content ($\phi_w$). (c) Schematic view of the realization of the patterns obtained by the combination of the electric and magnetic fields. (d) A scheme of the coupling between the electrokinetic driving and magnetostatic energy reduction. (e) Scheme of the top view of the microelectrode cell with electrode periodicity indicated. (f) Transmitted light microscopy image of the microelectrode cell filled with ferrofluid in absence of external fields and (g) a selection of images under different applied voltages and magnetic fields, demonstrating the five characteristic regimes.

## B. The effect of humidity

Accumulation of water in the cores of the reverse micelles is known to enhance spontaneously charging of the micelles and hence increase the electrical conductivity of micelle dispersions in nonpolar solvents [53,54]. To control water content in our samples, we equilibrated the samples against atmospheres of different known humidities [Fig. 2a]. The water content in the sample stabilized under a new humidity typically after a few days, resulting in mass fraction of water $\phi_w$ ranging from $\phi_w = 0.4\%$ to $\phi_w = 1.2\%$ depending on the gas phase humidity, as quantified using coulometric Karl-Fischer titration [Fig. 2b]. Increase in water content promoted formation of more intense nanoparticle gradients under electric driving [Fig. 2c, d], quantified from optical images as $-\ln(I/I_0)$ that is proportional to the change in the nanoparticle concentration (where $I$ is the transmitted light intensity in the non-equilibrium state and $I_0$ is the intensity in the equilibrium state) [26]. Similarly, the conductivity of the samples increased with the water content [Fig. 2e] as expected [55–57].

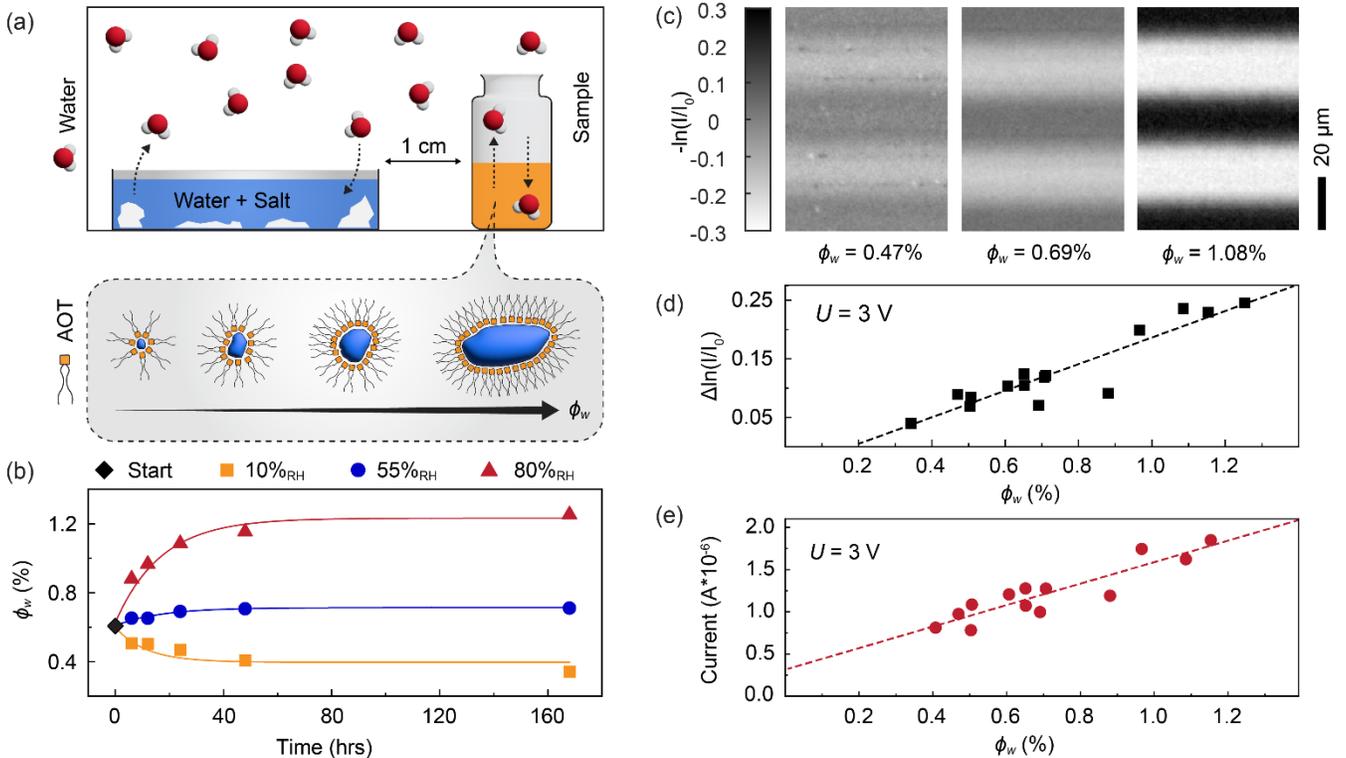

FIG. 2. Water content dependent nanoparticle gradients. (a) Scheme of equilibration of the sample against an atmosphere of fixed relative humidity (RH), and a schematic representation of the increase of the micelle size. (b) Volume fraction of water $\phi_w$ as a function of the incubation time. (c) Normalized

microscopy images taken of three samples with different $\phi_w$ driven to the gradient state ($U = 3.0$ V). (d) Peak-to-peak amplitude of $-\ln(I/I_0)$ as a function of $\phi_w$ ($U = 3.0$ V). (e) Electric current as a function of $\phi_w$ ($U = 3.0$ V).

## C. The elementary electrokinetic instability

The observed nanoparticle gradient becomes unstable when the applied was increased to ca. 5 V [Fig. 3]. Close to this threshold, the instability involves development of a periodic undulation pattern on the gradient with a typical periodicity $\lambda \approx 70$ μm [Fig. 3b]. We hypothesize that this undulation is driven and sustained by the electric field creating a tangential force on the perturbed space charge density existing on the diffuse interface, leading to a circulating microscopic flow [Fig. 3a,c]. This flow is further supported by microscopic observations [Video 1].

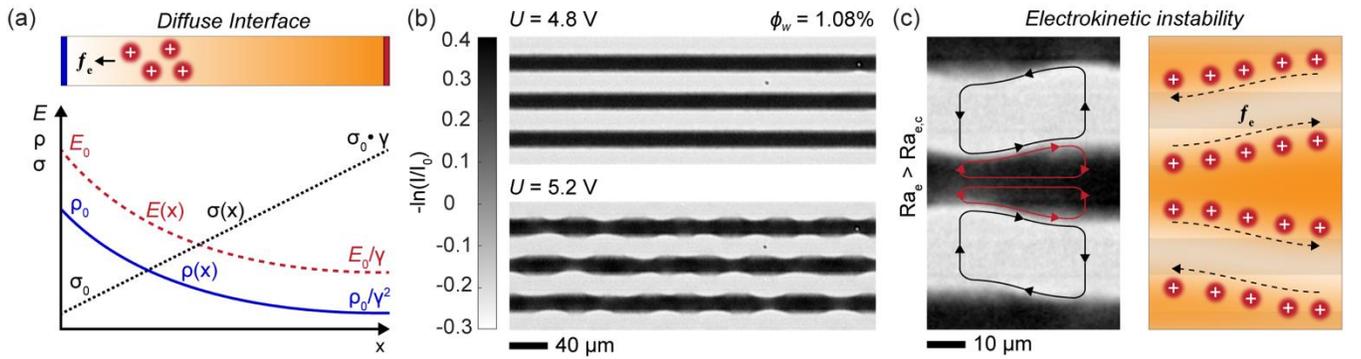

FIG. 3. Elementary electrokinetic instability. (a) Scheme of the nanoparticle gradient and a graph representing schematically the theoretical functional forms for electric field (dashed line) and space charge (solid line) under the assumption of linearly varying conductivity (short dash line) between two parallel plate electrodes ($\gamma = 1 + \Delta\sigma/\sigma_0$) [27]. (b) Normalized microscopy images of a single sample just below the electrokinetic instability ($U = 4.8$ V) and just above the threshold ($U = 5.2$ V). (c) Scheme of the hydrodynamic flows, electrostatic forces, and space charge distribution in the undulated state.

The observed instability can be rationalized by modifying previous theoretical considerations [32]. The link between the driving space charge density $\rho = \rho(r)$ and the underlying conductivity gradient $\sigma = \sigma(r)$ can be established from the Gauss's law, $\rho = \nabla \cdot (\varepsilon E)$, where $\varepsilon = \varepsilon(r)$ is the dielectric permittivity and $E = E(r)$ is the spatially varying electric field [32]. Using vector identity $\nabla \cdot (\psi A) = \psi \nabla \cdot A + A \cdot \nabla\psi$, the space charge density can be expressed as $\rho = \varepsilon \nabla \cdot E + E \cdot \nabla\varepsilon$, demonstrating that the charge density has two sources: divergence of electric field and gradient of permittivity. The divergence of the electric field is further linked to conductivity as $\nabla \cdot \boldsymbol{E} = -\frac{\boldsymbol{J}}{\sigma} \cdot \frac{\nabla \sigma}{\sigma}$ (that arises from the continuity equation $\nabla \cdot \boldsymbol{J} + \frac{\partial \rho}{\partial t} = 0$, where $\boldsymbol{J}$ is the current density under the assumption of steady state, $\frac{\partial \rho}{\partial t} = 0$, and ohmic conductivity, $\boldsymbol{J} = \sigma \boldsymbol{E}$). Hence, the space charge is given by

$$\rho = \varepsilon \nabla \cdot \boldsymbol{E} + \boldsymbol{E} \cdot \nabla \varepsilon \approx -\frac{\varepsilon \boldsymbol{J}}{\sigma} \cdot \frac{\nabla \sigma}{\sigma}, (1)$$

where the gradient of the dielectric constant was assumed to be small. Because the current density is approximately constant between the electrodes, the above expression predicts that the space charge tends to accumulate in regions where the gradient of the conductivity is large and the conductivity itself small [Fig. 3a].

In our experiments, the negatively charged nanoparticles accumulate near the positive electrodes – hence reducing conductivity near the negative electrode and promoting space charge there [Fig. 3a, c]. A force density $\boldsymbol{f}_e = -\rho \boldsymbol{E}$ creates a force towards the electrode of opposite sign. This force can hence induce destabilization of the gradient. It has been shown that the stability of a liquid layer with a constant conductivity gradient (increasing from $\sigma_0$ to $\sigma_0 + \Delta \sigma$ within the layer thickness $d$) in the direction of the electric field can be evaluated using a critical electric Rayleigh number

$$Ra_e = \frac{d^2 \varepsilon E_0^2}{\eta K_{eff}} \frac{\Delta \sigma}{\sigma_0}, (2)$$

where $E_0$ is the field strength on the lower conductivity boundary, $\eta$ is viscosity, and $K_{eff}$ is the diffusion coefficient of the charge carriers [28]. Assuming that the nanoparticles are the main charge carriers in our system, $K_{eff} = \frac{k_B T}{3\pi \eta D}$ (where $k_B$ is the Boltzmann constant, $T$ is temperature, and $D$ is hydrodynamic diameter of the nanoparticles). Furthermore, if one notes that the $E_0$ can be expressed as $(U/\ln(1 + \frac{\Delta \sigma}{\sigma_0}))/(\frac{\sigma_0}{\Delta \sigma}) = E_0 d$, then Eq. (2) becomes

$$Ra_e = 3\pi \frac{D\varepsilon U^2}{k_\mathrm{B} T} \frac{\left(\frac{\Delta\sigma}{\sigma_0}\right)^3}{\left(\ln\left(1+\frac{\Delta\sigma}{\sigma_0}\right)\right)^2}. \quad (3)$$

Plugging in values corresponding approximately to our system, $D \sim 12 \cdot 10^{-9}$ m, $\varepsilon \approx 2\varepsilon_0$, $U_c \sim 5$ V, $k_\mathrm{B}T = 4.14 \cdot 10^{-21}$ J and $\Delta\sigma/\sigma_0 = 10$ (order of magnitude estimate from IV measurements [25]), yields $Ra_e \sim 2 \cdot 10^6$, which is larger than the theoretically predicted critical $Ra_e \sim 1.5 \cdot 10^4$ at $\Delta\sigma/\sigma_0 = 10$ for the layer to become unstable [28]. This order of magnitude estimate suggests that the mechanism for the nanoparticle gradient for becoming unstable is indeed in the space charge density that experiences a force in the electric field.

### D. More complex electrokinetically driven spatiotemporal patterns

At higher applied voltages, the tangential flows on the undulated nanoparticle gradient grow quickly, leading to increasingly complex time-varying behaviors [Fig. 4]. The simplest of these behaviors is a dynamic pattern consisting of travelling waves [Fig. 4a, Video 2], wherein the convective flows spontaneously become asymmetric leading to the wave propagation along the electrodes. At higher water content and higher dissipation, this asymmetry becomes more pronounced and induces lateral accumulation of the nanoparticles and focused soliton-like fountains [Fig. 4b, Video 3]. At even stronger electric driving, the system enters a seemingly chaotic state with apparently random spatial and temporal changes of the nanoparticles concentration [Fig. 4c, Video 4]. Kymographic analysis shows that the travelling waves can exhibit well defined wavelengths at the length scale of tens of micrometers [Fig. 4a, d, Video 2] with translational speeds increasing linearly with the applied voltage [Fig. 4e]. Overall, the behavior can be summarized in a phase diagram [Fig. 4f], suggesting that higher water content and larger applied voltages promote the emergence of increasingly complicated and dynamic structures [Fig. 4g].

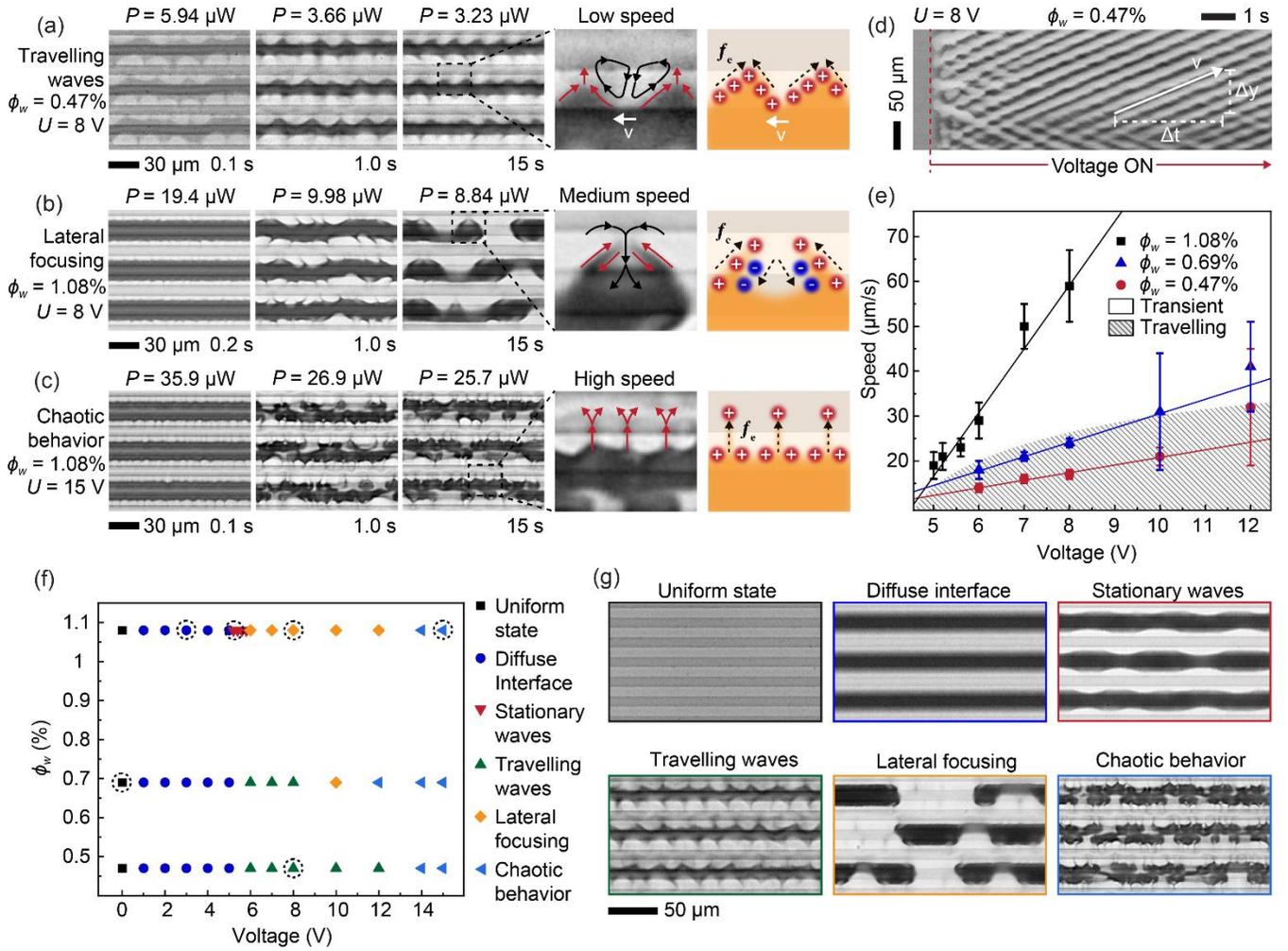

FIG. 4. Complex electrokinetically driven spatiotemporal patterns. (a-c) Time series of microscopy images showing the formation of three typical electrokinetic spatiotemporal patterns in the order of increasing dissipation: (a) travelling waves ($\phi_w = 0.47\%$, $U = 8$ V), (b) lateral focusing ($\phi_w = 1.08\%$, $U = 8$ V) and (c) seemingly chaotic behavior ($\phi_w = 1.08\%$, $U = 15$ V). (d) Kymograph generated from the time series in panel (a) by using the image intensity along the horizontal line in the middle of the gap between a positive and a negative electrode. (e) Plot of the speed of the moving waves as a function of applied voltage. Error bars indicate standard deviation of multiple measurements from different locations within the microelectrode cell. (f) Non-equilibrium phase diagram of different patterns and structures observed as a function of electric driving (x-axis) and $\phi_w$ (y-axis) and (f) representative microscopy images.

### E. Magnetic tuning of the patterns

Application of magnetic field over the electrokinetically patterned states [Fig 3,4] leads to a variety of new states [Fig. 5]. For example, application of a magnetic field perpendicular to a stationary undulated gradient leads to emergence of a finger pattern, with a drastically increased aspect ratio [Fig. 5a, Video 5]. On the other hand, a laterally focused state can be brought back to a laterally continuous state with much finer structures by application of an out-of-plane magnetic field [Fig. 5b, Video 6]. Even the apparently chaotic state can be tamed by application of an in-plane magnetic field along the electrodes, giving rise to a much more well-defined pattern [Fig. 5c, Video 7]. Application of an in-plane magnetic field perpendicular to the electrodes over a stationary wave pattern leads to another lateral focusing state [Fig. 5d, Video 8].

These magnetically driven tuning of the electrokinetic patterns can be understood to originate from reduction of the magnetostatic energy $U_m = -\frac{1}{2}\mu_0 \int_V \boldsymbol{M}(\boldsymbol{r}) \cdot \boldsymbol{H_0}(\boldsymbol{r}) dr^3$, where $\boldsymbol{M}$ is the local magnetization and $\boldsymbol{H_0}$ is the applied magnetic field [40]. Assuming linear isotropic medium, $\boldsymbol{M}(\boldsymbol{r}) = \chi(\boldsymbol{r})\boldsymbol{H}(\boldsymbol{r})$, where $\boldsymbol{H} = \boldsymbol{H}(\boldsymbol{r}) = \boldsymbol{H_0} + \boldsymbol{H_d}(\boldsymbol{r})$ is the total magnetic field that depends on the demagnetizing field $\boldsymbol{H_d}(\boldsymbol{r})$, allows rewriting the magnetostatic energy as

$$U_m = -\frac{1}{2}\mu_0 \int_V \chi(\boldsymbol{r}) \boldsymbol{H_0} \cdot \boldsymbol{H_0} dr^3 - \frac{1}{2}\mu_0 \int_V \chi(\boldsymbol{r}) \boldsymbol{H_d}(\boldsymbol{r}) \cdot \boldsymbol{H_0} dr^3 . \quad (4)$$

Because $\boldsymbol{H_0}$ is constant, Eq. (4) further simplifies to

$$U_m = -\frac{1}{2}\mu_0 \boldsymbol{H_0^2} \chi_{AVG} V - \frac{1}{2}\mu_0 \int_V \chi(\boldsymbol{r}) \boldsymbol{H_d}(\boldsymbol{r}) \cdot \boldsymbol{H_0} dr^3, \quad (5)$$

where $\chi_{AVG}$ is the volumetrically averaged susceptibility that is constant in the limit of non-interacting nanoparticles. The second term represent the effect of the demagnetizing field – therein a decrease in overall magnetostatic energy can be achieved as $\chi(\boldsymbol{r})$ and $\boldsymbol{H_d}(\boldsymbol{r})$ change with the redistribution of the nanoparticles. However, the redistribution of the nanoparticles also changes the spatially varying conductivity and hence the space charge density and electrokinetic forces. Thus, the steady-state nanoparticle distribution $n(\boldsymbol{r})$ needs to be solved self-consistently by considering both electrokinetic effects and energetic contributions from the magnetostatics [Fig. 1c]. This coupling is clearly manifested by a change in electric current and dissipation upon application of a magnetic field on the electrokinetically driven patterns [Fig. S2].

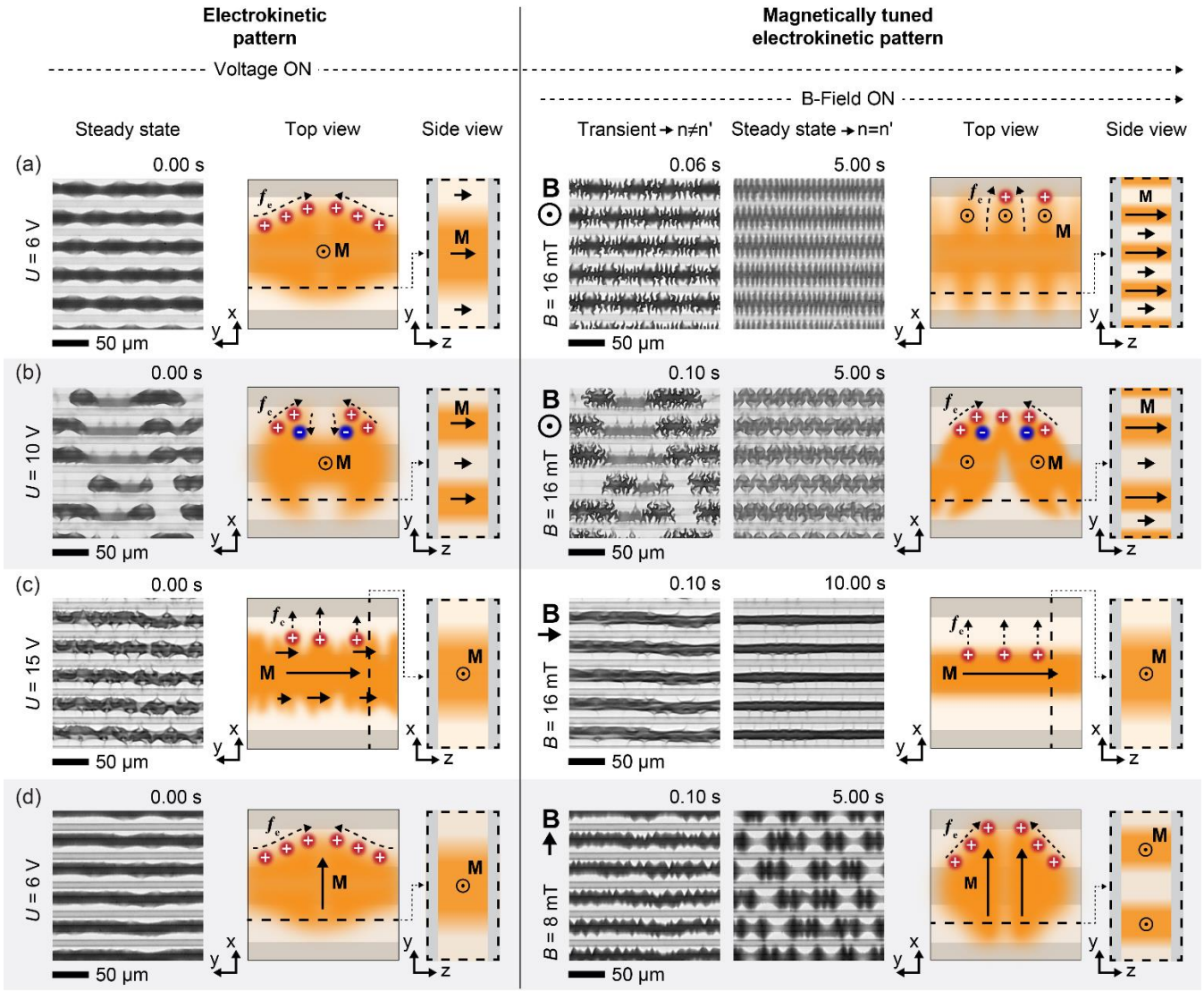

FIG. 5. Magnetically tuned electrokinetically driven non-equilibrium patterns. (a-d) From left to right, examples of experimentally observed electrokinetic patterns, corresponding schematic interpretations of the nanoparticle and charge distributions, two microscopy images of the magnetically tuned electrokinetic patterns and the hypothesized space charge and magnetization distributions.

## III. CONCLUSIONS

We have shown that electrically driven magnetic nanoparticle gradients in nonpolar solvents can undergo electrokinetic instability, leading to formation of numerous spatiotemporal patterns that can be further tuned by magnetic field via magnetostatic energy reduction. We have also experimentally demonstrated that increasing ambient humidity and water content in the samples promotes the nanoparticle gradient formation. We note that the complete theoretical understanding of the observed phenomena requires coupling and self-consistent solution of the electrokinetic and magnetostatic equations.

From materials science and technological perspectives, the ability to generate nanoparticle gradients in-situ using electrophoresis in contrast to typical systems utilizing microfluidic approaches [27,28,32,45] is appealing, as well as the diversity of the dynamic responses obtained at modest driving voltages (few volts) and in small magnetic fields (few milliteslas). Furthermore, as the magnetic nanoparticles and ferrofluids are used in various applications [58] to tune viscosity [59], local temperature [60], optical properties [61], and heat transfer [62], we foresee that the observed non-equilibrium patterns can become technologically relevant.

**Acknowledgements:**
This work was supported by ERC (803937), Academy of Finland (316219, 342116 and 342038), and ERC-2016-ADG-742829 DRIVEN. Work was carried out under the Academy of Finland Center of Excellence Program (2022-2029) in Life-Inspired Hybrid Materials LIBER (346112, 346108).

**Author Contributions:** FS, TC and CR synthesized the nanoparticles, prepared the samples, constructed the experimental setups, characterized the samples, and analyzed the data. FS and TC carried out the preliminary experiments and acquired the final microscopy data and electric field measurements. CR and TC wrote the first draft of the manuscript and CR and FS prepared the first draft of the figures. CR compiled the final figures and FS the supplementary videos. JVIT and CR developed the theoretical understanding. OI supervised TC and assisted in compiling the final manuscript. CR and JVIT wrote the final version of the manuscript. JVIT conceived the concepts and supervised the experiments and data analysis.

# Magnetically tunable electrokinetic instability and structuring of non-equilibrium nanoparticle gradients

F. Sohrabi,[†1] C. Rigoni,[†1] T. Cherian,[1] O. Ikkala,[1] J. V. I. Timonen,[1]*

[1]Department of Applied Physics, Aalto University School of Science,

Puumiehenkuja 2, 02150 Espoo, Finland

*Corresponding author; E-mail: jaakko.timonen@aalto.fi

[†] Equal contribution

A. **Material and methods**
B. **Supplementary figures**
C. **Supplementary videos**

## A. Material and methods

1. *Materials:* Iron (III) chloride hexahydrate (FeCl$_3 \cdot$ 6H$_2$O, $\geq$ 99%, Sigma-Aldrich), iron (II) sulfate heptahydrate (FeSO$_4 \cdot$ 7H$_2$O, $\geq$ 99%, ACS reagent, Sigma-Aldrich), ammonium hydroxide (NH$_4$OH, 28-30%, ACS reagent, Sigma-Aldrich), oleic acid (90% technical grade, Sigma-Aldrich), acetone ($\geq$ 99.8%, Fisher Scientific), toluene ($\geq$ 99.7%, ACS reagent, Sigma-Aldrich), *n*-dodecane (99%, anhydrous, Acros Organics), docusate sodium salt (AOT, $\geq$99%, anhydrous, Sigma-Aldrich), magnesium nitrate hexahydrate, (Mg(NO$_3$)$_2 \cdot$ 6H$_2$O, 99%, ACS reagent, Sigma-Aldrich), lithium chloride (LiCl, 99%, ACS reagent, Sigma-Aldrich), ammonium sulfate ((NH$_4$)$_2$SO$_4$, 99%, ACS reagent, Sigma-Aldrich), and microelectrode cells with interdigitated ITO electrodes (IPS10X10A040uNOPI, Instec) were used as received.

2. ***Synthesis and characterization of the stock dispersion of iron oxide nanoparticles in toluene:*** Synthesis and characterization of the iron oxide nanoparticles was done as previously [26] with small modifications. Briefly, the nanoparticles were synthesized using the Massart coprecipitation method [63] utilizing oleic acid as stabilizing surfactant and toluene as the final carrier fluid. The diameter of the nanoparticles was extracted from TEM images using image analysis [Fig. S1a, b] and the composition of the nanoparticles was verified using XRD [Fig. S1c] and Raman spectroscopy [Fig. S1d]. The mass fraction of the nanoparticles in the stock dispersion was determined by measuring the weight loss during evaporation of the carrier solvent using an analytical balance. The corresponding volume fraction was calculated by using the density of the ferrofluid. The results of the characterization are summarized in Table 1.

Table 1. Physical properties of the iron oxide nanoparticles and their stock dispersion in toluene.

| Measured property | Measured value |
|---|---|
| Mean nanoparticle core diameter | 8.5 ± 0.3 nm |
| Standard deviation of the nanoparticle core | 2.6 ± 0.1 nm |
| Density of the stock dispersion | 0.964 ± 0.001 g/ml |
| Mass fraction of nanoparticles in the stock | 14.6 ± 0.3 % |
| Volume fraction of nanoparticles in the stock | 3.6 ± 0.3 % |

3. ***Preparation and characterization of the stock electroferrofluid:*** Preparation and characterization of the electroferrofluid was done as previously [26] with small modifications. Briefly, the stock electroferrofluid was prepared by mixing 300 μl of the stock dispersion of iron oxide nanoparticles in toluene with 200 μl of a 150 mM solution of AOT in dodecane in a small glass vial. The vial was left open in a fume hood, under ambient temperature and humidity, for ca. 80 hours. At that point, FTIR measurement indicated that the toluene had fully evaporated [Fig. S1e]. The volume fraction of the nanoparticles was determined as for the stock ferrofluid. Magnetic properties of the stock electroferrofluid were measured using a vibrating sample magnetometer [Fig. S1f]. The results of the characterization of the stock electroferrofluid are summarized in Table 2.

Table 2. Physical properties of the stock electroferrofluid.

| Measured property | Measured value |
|---|---|
| Saturation magnetization | 8.3 ± 0.2 kA/m |
| Magnetic susceptibility | 0.243 ± 0.001 |
| Volume fraction of AOT | 5.8 ± 0.5 % |
| Volume fraction of nanoparticles | 5.1 ± 0.6 % |

4. ***Regulation of water content in the electroferrofluid:*** Small amounts of the stock electroferrofluid were equilibrated against various atmospheres with different RH levels in order to regulate the water content in the electroferrofluid. The controlled atmospheres were created inside plastic chambers (SICCO Mini desiccator cabinet) using saturated aqueous salt solutions: lithium chloride for 11% RH, magnesium nitrate hexahydrate for 55% RH, and ammonium sulfate for 80% RH. Stability of temperature and RH within each chamber was determined using a temperature and humidity data logger (Testo 174H). The temperature varied typically between 20°C and 22°C during the sample equilibration. The RH stayed within 2 percentage points from the nominal values.

5. ***Determination of water content in the electroferrofluids:*** The amount of water was determined using Coulometric Karl Fischer titration (Mettler Toledo C30S Compact KF Coulometer) with the oven method (Mettler Toledo InMotion KF Flex Oven Autosampler). Briefly, empty measurement vials were first dried in an oven at 80°C for 30 min. A small amount (ca. 100 μl) of the electroferrofluid was pipetted into a measurement vial and the mass of the fluid was determined using an analytical balance (Mettler Toledo XS204). The vial was sealed and heated to 130°C in the titrator oven to release the water that was subsequently quantified by the titrator. The mass fraction of water in the electroferrofluid was obtained by dividing the mass of titrated water by the mass of the electroferrofluid.

6. ***Imaging of the electroferrofluid under electric and magnetic fields:*** Imaging of the electroferrofluids was done was as previously [26]. Briefly, microelectrode cells were connected to metal wires using silver paste (SPI Supplies OK-SPI) and filled with the electroferrofluid using a micropipette (Eppendorf Research Plus). The electric field was created by applying a voltage, while simultaneously measuring the current, on the microelectrode cell using a high resistance meter (Keysight B2987A). The magnetic field was applied using a pair of electromagnet coils (GMW 11801523 and 11801524) connected to DC power supply (BK Precision 9205) as previously. Imaging was done using an optical microscope in the transmitted light mode.

## B. Supplementary figures

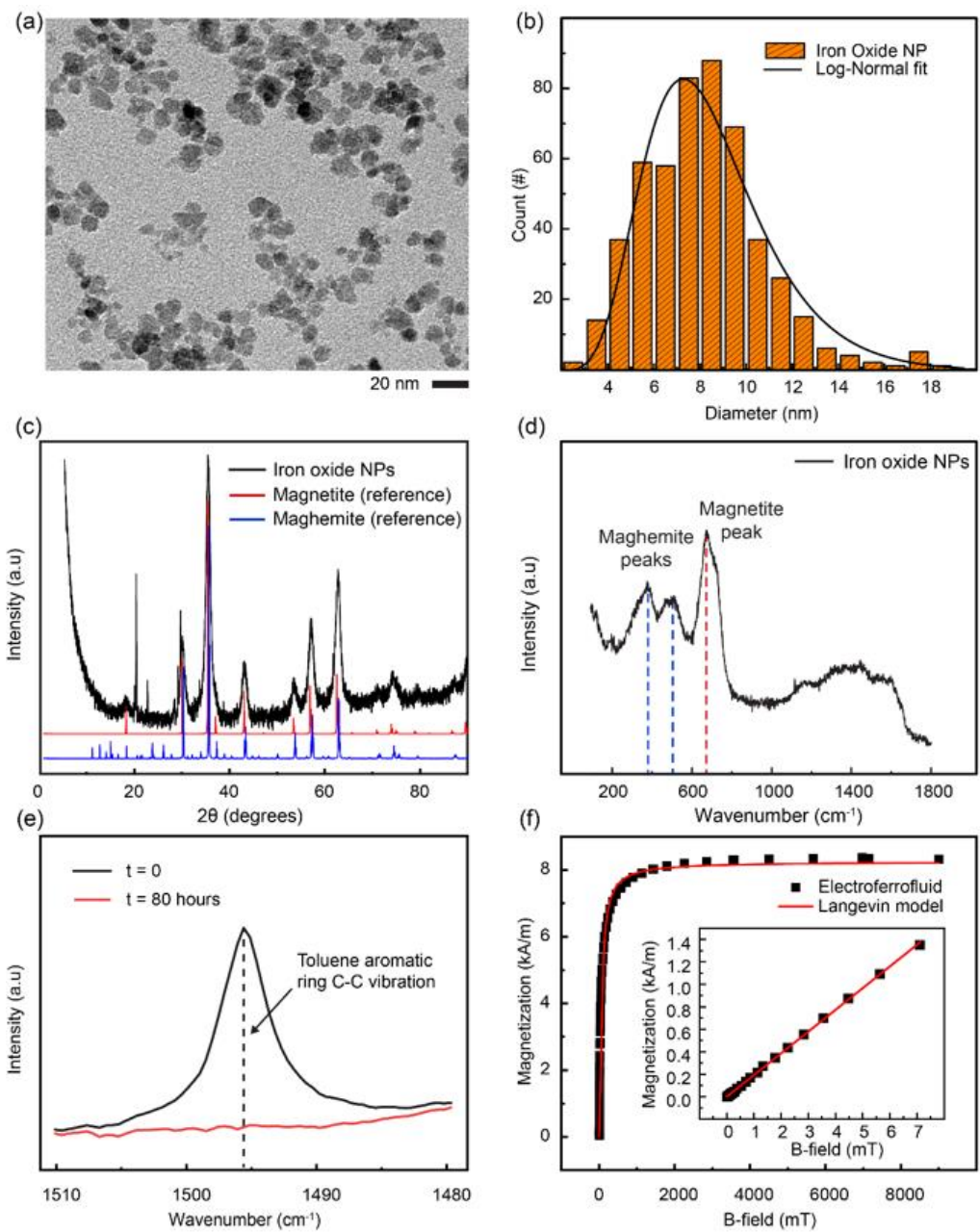

FIG. S1. Characterization of superparamagnetic iron oxide nanoparticles and the stock electroferrofluid. (a) A typical TEM image of the iron oxide nanoparticles stabilized with oleic acid. (b) A nanoparticle diameter histogram obtained from TEM images with a log-normal fit. (c) A powder XRD diffractogram of the iron oxide nanoparticles and references peaks. (d) A Raman spectrum of the iron oxide nanoparticles and references peaks of pure magnetite and maghemite. (e) FTIR spectra taken during the preparation of the electroferrofluid: immediately after preparing the mixture between iron oxide nanoparticles in toluene and 150 mM AOT in dodecane, and after 80 hours of evaporation. (f) A magnetization curve of the stock electroferrofluid and the best fit of Langevin model. The inset is a zoom of the data near the origin.

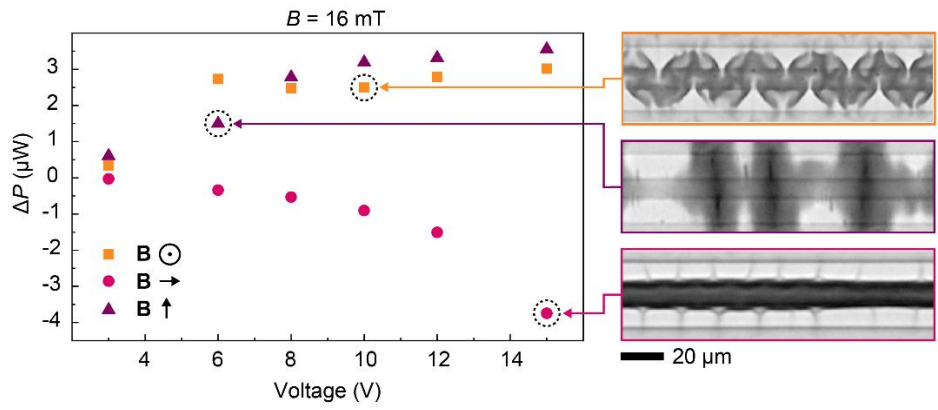

FIG. S2. Power dissipation change caused by application of magnetic field in magnetically tuned electrokinetic patterns.

## C. Supplementary videos

**Video 1.**

Stationary dissipative pattern formation (undulated gradient). Video shows a series of microscopy images when $U = 5.2$ V is applied for approximately 30 seconds to electroferrofluid with $\phi_w = 1.08\%$. Video was acquired at 250 fps and displayed in real time.

**Video 2**.

Travelling wave formation. Video shows a series of microscopy images when $U = 8$ V is applied for to an electroferrofluid with $\phi_w = 0.47\%$. Video was acquired at 250 fps and displayed in real time.

**Video 3.**

Lateral focusing. Video shows a series of microscopy images when $U = 8$ V is applied for approximately 30 seconds to an electroferrofluid with $\phi_w = 1.08\%$. Video was acquired at 250 fps and displayed in real time.

**Video 4.**

Seemingly chaotic behavior. Video shows a series of microscopy images when $U = 15$ V is applied for approximately 30 seconds to an electroferrofluid with $\phi_w = 1.08\%$. Video was acquired at 250 fps and displayed in real time.

**Video 5.**

Coupling between a stationary wave pattern driven by $U = 6$ V and an out-of-plane $B = 16$ mT in an electroferrofluid with $\phi_w = 1.08\%$. Video was acquired at 250 fps and displayed in real time.

**Video 6.**

Coupling between the lateral focusing state driven by $U = 10$ V, and an out-of-plane $B = 16$ mT in an electroferrofluid with $\phi_w = 1.08\%$. Video was acquired at 250 fps and displayed in real time.

**Video 7.**

Coupling between a chaotic state driven by $U = 15$ V, with an in-plane (along electrodes) $B = 16$ mT in an electroferrofluid with $\phi_w = 1.08\%$. Video was acquired at 250 fps and displayed in real time.

**Video 8.**

Coupling between a stationary wave pattern driven by $U = 6$ V and an in-plane (perpendicular to the electrodes) $B = 8$ mT in an electroferrofluid with $\phi_w = 1.08\%$. Video was acquired at 250 fps and displayed in real time.